%% file: main.tex
\documentclass[10pt,conference]{IEEEtran}
\IEEEoverridecommandlockouts
\usepackage{cite}
\usepackage{amsmath,amssymb,amsfonts}
\usepackage{algorithmic}
\usepackage{graphicx}
\usepackage{textcomp}
\usepackage{xcolor}

\usepackage{subfiles} % Best loaded last in the preamble
\usepackage{multirow}
\usepackage[ruled,linesnumbered]{algorithm2e}
\usepackage{listings}
\usepackage{subcaption}
\usepackage{color,xcolor}
\usepackage{color,xcolor,colortbl}
\usepackage{xspace}
\usepackage{enumitem}
\usepackage{graphicx}
\usepackage{xr}
\usepackage{url}

\usepackage{tcolorbox}

\usepackage{booktabs}
\usepackage{diagbox}
\usepackage{amsmath}
\usepackage{array}
\makeatletter
\newcommand{\thickhline}{%
    \noalign {\ifnum 0=`}\fi \hrule height 1pt
    \futurelet \reserved@a \@xhline
}
\newcolumntype{"}{@{\hskip\tabcolsep\vrule width 1pt\hskip\tabcolsep}}
\makeatother

\newtheorem{Def}{Definition}
\newcommand{\tool}{PILOT\xspace}
\newcommand{\eg}{\textit{e}.\textit{g}.\xspace}
\newcommand{\ie}{\textit{i}.\textit{e}.\xspace}
\newcommand{\et}{\textit{et} \textit{al}.}

\newcommand{\http}{\url{https://github.com/PILOT-VD-2023/PILOT}}

\def\BibTeX{{\rm B\kern-.05em{\sc i\kern-.025em b}\kern-.08em
    T\kern-.1667em\lower.7ex\hbox{E}\kern-.125emX}}
\begin{document}
\title{When Less is Enough: Positive and Unlabeled Learning Model for Vulnerability Detection}
% \title{Conference Paper Title*\\

% \and
% \and
% \and
% \and
% \and
\author{Anonymous Author(s)}
\author{\IEEEauthorblockN{Xin-Cheng Wen$^{1}$, Xinchen Wang$^{1}$, Cuiyun Gao$^{1\ast}$, Shaohua Wang$^{2}$, Yang Liu$^{3}$, Zhaoquan Gu$^{1}$}

\IEEEauthorblockA{$^1$ School of Computer Science and Technology, Harbin Institute of Technology, Shenzhen, China}

\IEEEauthorblockA{$^2$ Central University of Finance and Economics, China}

\IEEEauthorblockA{$^3$ School of Computer Science and Engineering, Nanyang Technological University, China}

\IEEEauthorblockA{xiamenwxc@foxmail.com, 200111115@stu.hit.edu.cn, davidshwang@ieee.org, \\ yangliu@ntu.edu.sg, 
\{gaocuiyun, guzhaoquan\}@hit.edu.cn}

\thanks{$^{\ast}$ Corresponding author. The author is also affiliated with Peng Cheng Laboratory.}}

\maketitle

\begin{abstract}

\label{sec:abstract}
\input{Sections/0_abstract}

\end{abstract}


\section{Introduction}
\label{sec:introduction}
\input{Sections/1_Introduction}

\section{Assumptions of PU Learning}
%\section{Background}
\label{sec:background}

\input{Sections/2_Background_Motivation}

\section{Proposed Framework}
\label{sec:architecture}

\input{Sections/3_Method}

\section{EXPERIMENTAL Setup}
\label{sec:evaluation}
\input{Sections/4_Evaluation}

\section{Experimental Results}
\label{sec:experimental_result}
\input{Sections/5_Experimental_Result}

\section{Discussion}
\label{sec:discussion}
\input{Sections/6_Discussion}

\section{Related Work}
\label{sec:related}
\input{Sections/7_Related_Work}

\section{Conclusion}
\label{sec:conclusion}
\input{Sections/8_Conclusion}

\section*{Acknowledgment}

This research is supported by National Key R\&D Program of China (No. 2022YFB3103900), National Natural Science Foundation of China under project (No. 62002084), Natural Science Foundation of Guangdong Province (Project No. 2023A1515011959), Shenzhen Basic Research (General Project No. JCYJ20220531095214031), and the Major Key Project of PCL (Grant No. PCL2022A03).

\bibliographystyle{IEEEtran} 

\bibliography{IEEEabrv, Citation}

\end{document}

%% file: Sections/0_abstract.tex
% In recent years, automated code vulnerability detection has garnered increasing attention.
% The advancements in machine learning (ML) and deep learning (DL) methods have enabled models to implicitly learn vulnerable code patterns. Although existing methods have achieved some success, one challenge is that existing vulnerability detection (VD) models require large amounts of labeled data to facilitate learning. 
% To address the issue of missing labels, current methods usually rely on statically analyzed code or human-generated submissions, making it difficult to determine the accuracy of these labels.
% In this paper, we propose \tool, a novel positive and unlabeled learning model for vulnerability detection that emphasizes handling data-missing situations. \tool aims to address the shortcomings of vulnerability detection in the presence of missing data by leveraging known vulnerability samples to learn vulnerability patterns and then utilizing unlabeled data to further optimize the model for detecting unknown vulnerabilities. We conducted several experiments to evaluate \tool using real-world vulnerability datasets. \wxc{xx}
% In recent years, automated code vulnerability detection has garnered increasing attention. The advancements in machine learning (ML) and deep learning (DL) methods have enabled models to implicitly learn vulnerable code patterns. 
Automated code vulnerability detection has gained increasing attention in recent years. The deep learning (DL)-based methods, which implicitly learn vulnerable code patterns, have proven effective in vulnerability detection. The performance of DL-based methods usually relies on the quantity and quality of labeled data. However, the current labeled data are generally automatically collected, such as crawled from human-generated commits, making it hard to ensure the quality of the labels. Prior studies have demonstrated that the non-vulnerable code (i.e., \textit{negative labels}) tends to be unreliable in commonly-used datasets, while vulnerable code (i.e., \textit{positive labels}) is more determined. Considering the large numbers of unlabeled data in practice, it is necessary and worth exploring to leverage the positive data and large numbers of unlabeled data for more accurate vulnerability detection. 
%there are still limited by large amounts of labeled data to facilitate learning. 
%In addition, 
% The current labeled data usually rely on statically analyzed tools or human-generated commits, which it leads difficult to get high-quality labels. 
% Previous works have also demonstrated that non-vulnerable (negative) labels for source
% functions are unreliable in the real-world scenario. 
% Therefore, the main challenge is to mitigate the problem of poor-quality labels by using the vulnerable (positive) labeled data and a large number of unlabeled data for vulnerability detection.

%Besides, they have difficulty identifying vulnerable software that contains only a small number of labels.

% To mitigate the above issues, 
In this paper, we focus on the Positive and Unlabeled (PU) learning problem for vulnerability detection and propose a novel model named \textbf{\tool}, i.e., \textbf{P}osit\textbf{I}ve and unlabeled \textbf{L}earning m\textbf{O}del for vulnerability de\textbf{T}ection. 
\tool only learns from positive and unlabeled data for vulnerability detection.
It mainly contains two modules: (1) A distance-aware label selection module, aiming at generating pseudo-labels for selected unlabeled data, which involves the inter-class distance prototype and progressive fine-tuning;
% aims to address the shortcomings of vulnerability detection in the presence of missing labels 
% by leveraging positive data to learn vulnerability patterns and then utilizing pseudo-labels to further detect unlabeled vulnerabilities. 
(2) A mixed-supervision representation learning module to further alleviate the influence of noise and enhance the discrimination of representations. 
Extensive experiments in vulnerability detection are conducted to evaluate the effectiveness of \tool based on real-world vulnerability datasets. The experimental results show that \tool outperforms the popular weakly supervised methods by 2.78\%-18.93\% in the PU learning setting. Compared with the state-of-the-art methods, \tool also improves the performance of 1.34\%-12.46\% in F1 {score} metrics in the supervised setting. 
In addition, \tool can identify 23 mislabeled from the FFMPeg+Qemu dataset in the PU learning setting based on manual checking.

% \ly{first point above the sentence is too long, and PU/PUVD is not explained before using.}

%% file: Sections/1_Introduction.tex
Software vulnerabilities typically refer to specific flaws or oversights within software components that enable attackers to disrupt a computer system or program. In 2022, Trellixd's team uncovered a security vulnerability in Python's tarfile module, known as CVE-2007-4559~\cite{mckee2023trellix}. This vulnerability had been present in approximately 350,000 open-source projects, potentially creating a significant security risk due to it going unnoticed for so long. Consequently, researchers seek to improve approaches for software vulnerability detection in order to safeguard computer systems and programs from potential attacks.

\begin{figure}[t]
	\centering
	\includegraphics[width=0.48\textwidth]{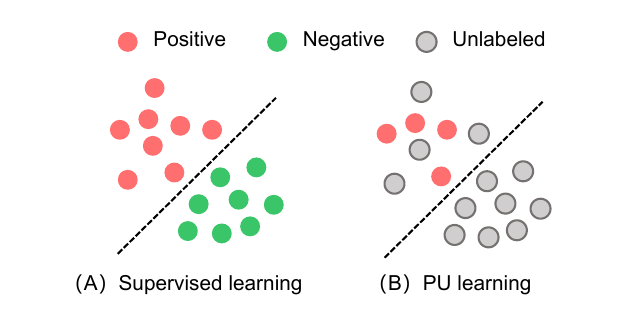}
    \caption{(A) illustrates supervised learning model is trained on a set of positive and negative samples. (B) represents Positive and Unlabeled (PU) learning models on the training set which only contains few
    % some 
    labeled positive and some
    % many 
    unlabeled samples. The red, green, and grey circles denote positive, negative, and unlabelled samples, respectively.}
	\label{fig:chart}
\end{figure}

In recent years, with the increase in the number of software vulnerabilities~\cite{DBLP:journals/pieee/LinWHZX20}, more and more researchers have been using automated methods for software vulnerability detection. Generally, existing software vulnerability detection methods can be divided into two categories: program analysis (PA)-based methods~\cite{FLAWFINDER, INFER, CHECKMARX} and learning-based methods~\cite{nguyen2021regvd, IVDETECT, linevd, cao2022mvd,DBLP:conf/icse/GaoGWSLY23/cream}. PA-based methods mainly include static analysis~\cite{CLANG, DBLP:conf/acsac/ViegaBKM00/2000ITS4/, DBLP:conf/cc/SuiX16/svf}, dynamic analysis~\cite{fuzz, argos}, and symbolic execution~\cite{symbolic}, among others. These methods usually utilize expert knowledge to manually extract features. They primarily focus on specific types of vulnerabilities, \eg buffer overflow~\cite{DBLP:conf/uss/LarochelleE01/bufferoverflow} and SQL injection~\cite{DBLP:conf/acns/BoydK04/SQL}, etc. 

In comparison, learning-based methods can detect a broader range of software vulnerability types~\cite{DBLP:conf/issta/ChengZ0S22}, such as various vulnerability in libraries and API calls. Learning-based methods mainly include sequence-based~\cite{sysevr, vuldeepecker, russell} and graph-based approaches~\cite{IVDETECT, devign, reveal}, both of which require a large amount of annotated data for training. For instance, VulDeePecker~\cite{vuldeepecker} and SySeVR~\cite{sysevr} treat source code as sequences and extract code gadgets from the source code and use a bidirectional Long Short-Term Memory (LSTM) network for vulnerability detection.
% , to detect vulnerabilities from a vast amount of supervised data. 
Devign~\cite{devign} and Reveal~\cite{reveal} extract various graph structures (such as Data Flow Graphs~\cite{Dataflow}, and Control Flow Graphs~\cite{cfg}) from the code and leverage Gated Graph Neural Networks to detect vulnerabilities within the code. 

Although learning-based methods have made significant progress in software vulnerability detection, %these methods still have limitations.
the performance of these methods is still limited due to the lack of high-quality labeled data. Specifically, these methods suffer from the following limitations: 
{\bf {(1) Lack of labeled data.} }
One major limitation is that existing vulnerability detection methods require a large amount of labeled positive and negative samples. % in Figure~\ref{fig:chart}(A). 
%However, due to human and resource constraints, there is often insufficient manual labeling. 
% The number of software vulnerabilities is large and continues to grow every year. In 2021, 20,149 software vulnerabilities were reported on the National Vulnerability Database (NVD)~\cite{DBLP:conf/compsac/NazimFSKMSW22/21nvd}, which increased by 41.26\% to 26,448 reported vulnerabilities~\cite{nvd}.
%. In 2022, this number increased by 41.26\% to 26,448 reported vulnerabilities.~\cite{nvd} 
However, manual code review requires expert knowledge~\cite{DBLP:conf/icsm/HanLXLF17/know} and is time-consuming~\cite {DBLP:conf/pldi/JovanovicKK06/time}, resulting in the lack of high-quality labeled data.
% \wxc{it is challenging to manually label code examples of each vulnerability}. %\dw{here, I guess it is challenging to manually label code examples of each vulnerability, not the vulnerability}
{\bf (2) Inaccurate labeled data.} The accuracy of the labels in vulnerability detection
% labeling existing vulnerabilities 
is also a major challenge. 
% Previous works have demonstrated that non-vulnerable labels for source functions are unreliable in real-world datasets as there is no ground truth label source for this class~\cite{DBLP:journals/tse/CroftXB23/311, DBLP:conf/msr/CroftBC22/312}. 
The
% most 
commonly-used labeling methods crawl the labels from public commit~\cite{DBLP:conf/msr/FanL0N20/fan} or rely on 
% involve 
static analyzers~\cite{DBLP:conf/icse/ZhengPLBEYLMS21/D2A},
% or manual commits \yun{[such labeling method if involving manual checking is good enough?]}, 
which are not foolproof and prone to introducing inaccurate labels.
% make it challenging to ascertain label accuracy. 
Croft \et's~\cite{DBLP:journals/corr/abs-2301-05456/dataquality} research has also identified the presence of noisy data. For example, only 80\% of the labels in the FFMPeg+Qemu~\cite{devign} dataset are reported as accurate.
% which has only 80\% accuracy \yun{[in [xxx]]}. 
% In addition, 
The prior research has also demonstrated that the non-vulnerable
% (negative) 
labels (i.e., \textit{negative labels}) are low in quality,
% for source code
while the vulnerable labels (i.e., \textit{positive labels}) are more reliable
% due to the lack of a ground truth label source~
\cite{DBLP:journals/tse/CroftXB23/311, DBLP:conf/msr/CroftBC22/312}. 
% Therefore, 
Considering the large numbers of unlabeled data in practice, it is critical to use the higher-quality positive labels and the unlabeled labels for vulnerability detection.
% vulnerability 
% % (positive) 
% samples and a large number of unlabeled samples for vulnerability detection.
 %For instance, the dataset collected by Fan \et~\cite{DBLP:conf/msr/FanL0N20/fan} only captures commits before or after and subsequently identifies vulnerable or not from those commits.
%However, the accuracy of the labels generated through this method cannot be guaranteed. 

% The existing labeling methods are mostly annotated using static analyzers or manual commits, making it difficult to determine the accuracy of the labels. For example, the Fan \et~\cite{DBLP:conf/msr/FanL0N20/fan} dataset collects data only on commits before or after and then extracted vulnerable or non-vulnerable functions samples from the commits. 
% The labels' accuracy obtained by this method cannot be guaranteed. The research by Croft \et~\cite{DBLP:journals/corr/abs-2301-05456/dataquality} elucidates the presence of noisy data in existing datasets, such as FFMPeg+Qemu~\cite{devign} dataset only has 80\% accuracy. 
% While the most advanced methods employ various state-of-the-art techniques for vulnerability detection, they overlook the learning capability of the model itself and the robustness of the labels. 
% The model's ability to detect vulnerabilities can be misled by mislabelled data.

To address the challenges above, we propose a \textbf{P}osit\textbf{I}ve and unlabeled \textbf{L}earning m\textbf{O}del for vulnerability de\textbf{T}ection, called \textbf{\tool}.  \tool mainly contains two modules: (1) A distance-aware label selection module, which generates pseudo-labels for selecting high-quality unlabeled data. It consists of the inter-class distance prototype and progressive fine-tuning; (2) A mixed-supervision representation learning module to further alleviate the influence of noise and enhance the discrimination of the vulnerability representation. As shown in Figure~\ref{fig:chart}, different from the supervised learning setting, the positive and unlabeled (PU) learning setting (i.e., PU setting) only requires a few labeled positive samples and some unlabeled samples for training.
% compared with the supervised learning method, we propose a Positive and Unlabeled (PU) learning problem (\ie PU setting) that only contains some labeled positive and many unlabeled samples on the training set.

% To evaluate \tool{}, we use three widely-studied benchmark datasets: {FFMPeg+Qemu~\cite{devign}, Reveal\cite{reveal}, and Fan et al.~\cite{fan}} and two settings: {PU and supervised learning} in software vulnerability detection. We compare \tool{} with four PU learning and four existing software vulnerability detection methods.

% In this study, 
We evaluate the effectiveness of \tool{} for detecting software vulnerabilities under
% using 
two settings: PU and supervised settings. Three popular benchmark datasets are adopted for the evaluation, including
% To assess its performance, we use three widely-studied benchmark datasets: 
FFMPeg+Qemu~\cite{devign}, Reveal\cite{reveal}, and Fan et al.~\cite{DBLP:conf/msr/FanL0N20/fan}. We compare \tool{} with four commonly used weakly supervised methods in the PU setting and five existing software vulnerability detection methods in the supervised setting. The results demonstrate that \tool outperforms all the baseline methods with respect to
% in 
the F1 score metric.
In particular, \tool{} achieves 2.78\%, 18.44\%, and 18.93\% in the PU setting on the three datasets, respectively, with relative improvements at 1.34\%, 12.46\%, and 3.00\% absolute improvement in the supervised setting.
In addition, \tool identifies
% can identify 
23 mislabeled samples from the training set of the FFMPeg+Qemu dataset
% in the PU learning setting 
based on manual checking, verifying the inaccurate label issue of existing labels.

In summary, the major contributions of this paper are summarized as follows:

\begin{enumerate}

\item We are the first to focus on the positive and unlabeled learning problem for software vulnerability detection. 
% We observe that \tool is
% % significantly 
% more effective in addressing the lack of high-quality labels situation compared to widely adopted baselines.

\item We propose \tool, a novel vulnerability detection framework under the PU setting. \tool involves a distance-aware label selection module for providing pseudo-label and a mixed-supervision representation learning module for alleviating the influence of noisy labels in
% influence for 
vulnerability detection.
% utilizes the enhanced graph representation learning method to learn the representations of simplified code structure graph
% for more accurate vulnerability detection.

\item We perform an evaluation of \tool on two settings and three public benchmark datasets, and the results demonstrate the effectiveness of \tool in software vulnerability detection.
\end{enumerate}

The remaining sections of this paper are organized as follows. Section~\ref{sec:background} introduces the assumptions of positive and unlabeled learning. Section~\ref{sec:architecture} presents the architecture of \tool, which includes two modules: a distance-aware label selection module and a mixed-supervision representation learning module. Section~\ref{sec:evaluation} describes the experimental setup, including datasets, baselines, and experimental settings. 
Section~\ref{sec:experimental_result} presents the experimental results and analysis.
Section~\ref{sec:discussion} discusses why \tool can effectively detect code vulnerability and the threats to validity.  Section~\ref{sec:conclusion} concludes the paper.

%% file: Sections/2_Background_Motivation.tex
\begin{figure*}[t]
	\centering
	\includegraphics[width=0.95\textwidth]{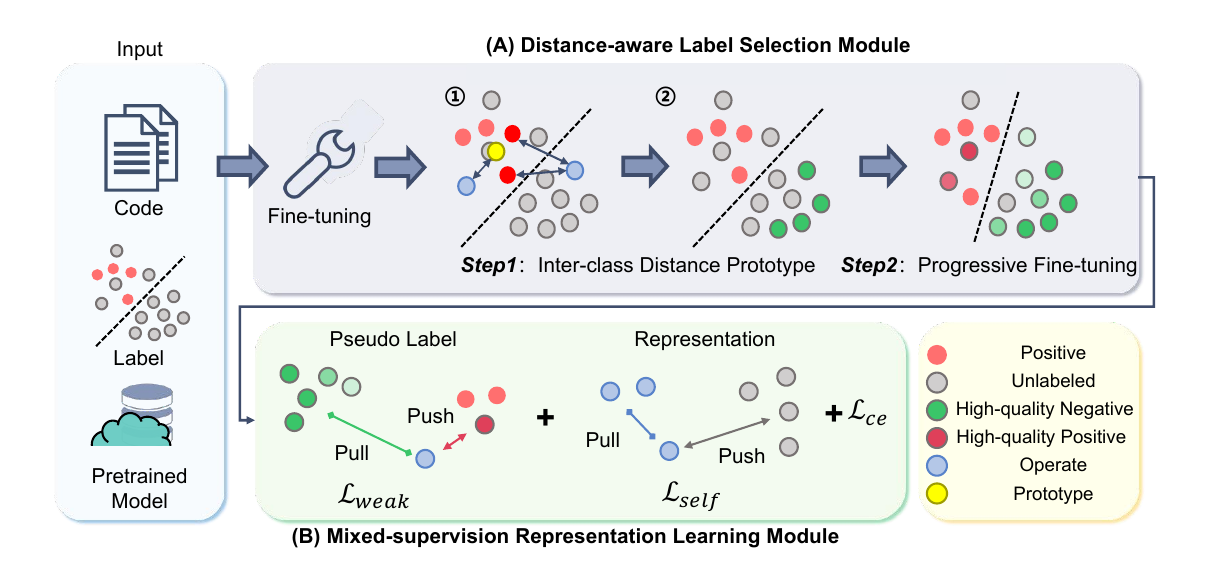}
    \caption{The architecture of \tool, which mainly contains two modules: (A) a distance-aware label selection module, and (B) a mixed-supervision representation learning
module. The different colored circles denote samples under different labels. The same color scheme denotes the same role. Different shades denote the order of labeling, with darker colors denoting earlier labeling.}
\label{fig:architecture}
\end{figure*}

%\subsection{Pre-trained Models of Code}

Positive and Unlabeled (PU) learning setting~\cite{DBLP:journals/ml/BekkerD20/pusurvey, DBLP:conf/cvpr/ColeALPMJ21/PU1,DBLP:conf/iccv/PantazisBJA21/PU2} is a weakly supervised classification setup where only PU samples are used for training. It does not require fully supervised data to obtain the same performance as supervised data. In this section, we introduce the assumptions
% and background settings 
of PU learning, including data assumption and label assumption.

\subsection{Data Assumption}
\label{sec:dataAssumption}
% In this paper, 
The PU data in this paper originate from a single training-set scenario~\cite{DBLP:conf/kdd/ElkanN08/singletraining}, meaning that the data come from one single training set.  
In PU settings, a fraction $c$ from the positive samples is selected to be labeled, following the dataset having a fraction $\alpha c$ of labeled samples. Specifically, the probability density functions of the ground truth distribution $f(x)$ is following:
\begin{align} 
\mathbf {x}&\sim f(x) \nonumber \\&\sim \alpha f_+(x)+(1-\alpha ) f_-(x) \nonumber \\&\sim \alpha c f_l(x) + (1-\alpha c) f_u(x). 
\end{align}
where $\alpha$ is the fraction of the positive samples in the dataset, $f_+(x)$, $f_-(x)$, $f_l(x)$ and $f_u(x)$ denote the probability density functions of the positive, negative, labeled and unlabeled samples, respectively.
In addition, PU learning is required to comply with the separability~\cite{DBLP:conf/icml/LiuLYL02/sepe1, DBLP:journals/jmlr/BlanchardLS10/sepe2} and smoothness~\cite{DBLP:conf/ijcai/LiL03/smoo1, DBLP:conf/esann/PelckmansS09/smoo2} assumptions.
\begin{Def}[Separability]
\label{Separability}
In the hypothesis space, there exists a function that can map positive samples to values greater than or equal to a threshold $\tau$, and negative samples to values below $\tau$.
\end{Def}

Under the separability assumption, the optimal classifier can classify all labeled samples as positive and a part of dissimilar unlabeled samples as negative. Considering that the optimal classifier is hard to be obtained, one commonly-used approach is to vary the threshold value $\tau$ and select a subset of samples to further train the classifier~\cite{DBLP:journals/ml/BekkerD20/pusurvey}.
% \wxc{Improving the quality of classification is often challenging. One effective approach is to vary the threshold value and select a subset of samples.}

%This is often difficult to achieve, one effective approach is to vary the threshold value and select a subset of samples, and we can improve the quality of the labeled sample\yun{[the grammar is wrong..]}.

\begin{Def}[Smoothness]
\label{Smoothness}
If the representations of two instances $x_1$ and $x_2$ are similar in the hypothesis space,
% then 
the probabilities $Pr(y=1|x_1)$ and $Pr(y=1|x_2)$ will also be similar.
\end{Def}

The smoothness assumption allows identifying high-quality negative samples as those that are far from all the positive samples.
% labeled examples. 
This can be done by using different similarity (or distance) learning measures. 
%In this paper, our proposed PU setting and conduct experiments are based on the smoothness assumptions.

% we use distance-aware label selection module (Section~\ref{prototype_section}) and mixed-supervision representation learning module (Section~\ref{metric_section}) methods to alleviate the influence of noise and enhance the discrimination of the vulnerability representation.

\subsection{Label Assumption}
\label{sec:labelAssumption}
% An 
Another important assumption of PU learning is 
% how we make assumptions 
about the labeling mechanism. It pertains to the selection process for instances labeled as vulnerable (\ie positive) in the experimental setup. In this paper, we choose the basic assumption for most weakly supervised methods~\cite{DBLP:conf/cvpr/ColeALPMJ21/PU1, DBLP:conf/iccv/PantazisBJA21/PU2, DBLP:journals/tnn/GongLYT19/PU3} in PU setting, called the Selected Completely At Random (SCAR) assumption~\cite{DBLP:conf/kdd/ElkanN08/singletraining}:

\begin{Def}[Selected Completely At Random]
\label{SAR}
Labeled samples are selected completely at random, independent of their representations, from the positive samples distribution $f_+(x)$. 
\begin{align} 
e(x) = \Pr (s=1|x,y=1)=\Pr (s=1|y=1)=c. \end{align}
\end{Def}

% In the SCAR assumption, each sample will have a propensity score $e(x)$, \dw{that} is the probability for selecting a positive example is constant and equal to the label frequency $c$ (\dw{this sentence has a little bit grammar issue}). 

In the SCAR assumption, the probability of selecting a positive sample is constant and equal to the label frequency $c$. Each sample will have a propensity score $e(x)$ based on the label frequency $c$.
%It is motivated by real-world scenarios, where 
The vulnerabilities observed by programmers are usually independent of the code represention~\cite{DBLP:conf/vizsec/AssalCB16/repre}. One example of a difficult-to-identify vulnerability is CWE-369 (Divide By Zero)~\cite{CWEID369}, in which
% typically 
the vulnerable
% vulnerability 
statements present
% represent 
only a small fraction in the source code. 
% It poses challenges for
Such vulnerabilities are hard to be identified by representation-based vulnerability detection methods~\cite{cheng2022bug}.
% similarity-based identification using code representation.
% Employing the probabilistic gap PU assumption~\cite{DBLP:journals/corr/abs-1808-02180} to select labeled samples based on code representation similarity may fail to identify this type of vulnerability. 
The evaluation of \tool in this paper is
% are 
based on the SCAR assumption.
% , and we have sampled repeatedly and replicated experiments to minimize errors. 

%(\dw{I get the point why use SCAR though, but I dont understand why the above example supports the assumption? why short trigger path and difficult to identify are relevant for choosing SCAR? perhaps revise here.})

%% file: Sections/3_Method.tex
In this section, we formulate the positive and unlabeled learning setting for vulnerability detection and then describe the overall framework of \tool.
As shown in Figure~\ref{fig:architecture}, \tool consists of two main modules: (1) a distance-aware label selection module for providing high-quality pseudo-labels (2) a mixed-supervision representation learning module to further alleviate the effects of noise and enhance the discriminative power of the vulnerability representation.

\subsection{Problem Formulation}

%The problem of positive and unlabeled (PU) learning for vulnerability detection is formulated below.
%We represent a positive and unlabeled (PU) learning task for vulnerability detection. 
A positive and unlabeled (PU) setting for vulnerability detection is 
%presented with the same goal as the general supervised vulnerability detection task: 
to train a binary classifier in the single-training-set scenario~\cite{DBLP:conf/kdd/ElkanN08/singletraining}, which can distinguish whether a sample is vulnerable or not based on the input source code.
In the PU learning setting,
only part of the vulnerable samples in the training data are labeled as positive and none of the non-vulnerable ones are labeled, as introduced in Section~\ref{sec:introduction}. 

We denote the PU dataset collected from the real world scenario 
%with training instances collected from the real world 
%A code function sample is denoted 
as $((x_{i}, y_{i}, l_{i})|x_{i} \in \mathcal{X}, y_{i} \in \mathcal{Y},  l_{i} \in \mathcal{C}, i\in\left\{ {1, 2,..., n} \right\} )$,  where $\mathcal{X}$ denotes the set of functions in the raw source code, $\mathcal{Y} = \left\{0,1\right\}$ denotes the class of code function (vulnerable or not), $\mathcal{C} = \left\{0,1\right\} $ denotes a binary variable representing whether the sample is selected to be labeled, $n$ is the number of code function in the dataset. In the PU setting of vulnerability detection, the class of sample $y$ is not observed, but the representation of source code
%\yun{[??]} 
can be derived from the label $c$.

The source code labeled with $c = 1$ indicates that it
% then this sample 
belongs to the vulnerable function (positive label):
% class): 
\begin{equation}
Pr(y=1 | c=1)=1
\end{equation}
% When
For the source code samples
% is 
unlabeled $c=0$, they can belong to vulnerable or not. 

Finally, \tool learns a mapping from $\mathcal{X}$ to $\mathcal{Y}$, $f: x_{i} \mapsto y_{i}$ to predict whether a code function is vulnerable or not. The prediction function $f$ is learned below:
\begin{equation}
\label{f}
min\sum_{i=1}^{n}\mathcal{L}\left(f\left(x_{i}, \hat{y_{i}}|\left\{x_{i}\right\}\right)\right)
\end{equation}
where $\mathcal{L}(\cdot)$ is the loss function, $\hat{y_{i}}$ is the pseudo label we extract from the input source code $x_{i}$.

\subsection{Distance-aware Label Selection Module}
\label{prototype_section}

In this section, we elaborate on the proposed distance-aware label selection module. It aims to identify High-quality Negative (HN) samples, which can provide pseudo-labeling of unlabeled samples. The module contains
% We propose 
two components, i.e., the inter-class distance prototype learning component and
% followed by 
progressive fine-tuning component.
In the following, we first elaborate on the two components and then provide an algorithm for the overall process.

\subsubsection{Inter-class Distance Prototype}

The inter-class distance prototype learning component
% method 
aims at identifying high-quality pseudo-labels in unlabeled samples. Labeled positive samples typically represent a small percentage of all training samples. Unlabeled samples may contain both positive and negative samples. Considering the diversity of vulnerabilities~\cite{DBLP:journals/tdsc/ZouWXLJ21},
%\yun{[cite]}, 
we select the corresponding positive prototype for each unlabeled sample based on the smoothness assumption (Section\ref{sec:dataAssumption}) and inter-class distance.
%.
%In contrast to previous approaches \yun{[???]}, 
To select HN samples, we leverage the distance between unlabeled samples and all labeled vulnerability samples. 
% We capture vulnerability patterns by considering the relationship between unlabeled samples and all labeled vulnerability samples, 
Only those unlabeled samples that exhibit 
%\yun{the most}
% significant 
the maximum distance difference is selected as HN samples.
% Only those unlabeled examples that exhibit significant differences from prototype examples are selected as reliable negative examples.

We use the CodeBERT~\cite{DBLP:conf/emnlp/FengGTDFGS0LJZ20/codebert} architecture to initialize the representations of code in the PU dataset $((x_{i}, l_{i})|x_{i} \in \mathcal{X}, l_{i} \in \mathcal{C}, i\in\left\{ {1, 2,..., n} \right\} )$.
%and leverage the initial weights pre-trained by Feng \et to do \wxc{initial} fine-tuning. 
%It enlarges the inter-class distance between positive and unlabeled samples and 
We generate the sequence vector $x^{0}$ for each source code function to capture both the global and local information, which is calculated as follows:

\begin{equation}
\label{cls}
x^{0} = H^{0}[CLS] +  H^{L}[CLS]
\end{equation}
where $H^{0}$ and $H^{L}$ denote the first and last layer embedding of CodeBERT~\cite{DBLP:conf/emnlp/FengGTDFGS0LJZ20/codebert}, respectively. CLS is a special token and is often used as the representation of the sequence~\cite{DBLP:conf/emnlp/LuHXKMDBLO21/CLS}.

To seek the inter-class distance prototype,
%In measuring the distance between samples, 
we use the positive samples close to the unlabeled samples. For each unlabeled sample $i$, the distance $D^{p}_{ij}$ to all positive samples $j$ is calculated. The top $k$ nearest samples are then selected, and the prototype of sample $i$ is the mean vector of the top $k$ samples. The sum and mean of the prototype distance for sample $i$ are calculated as $D^{p}_{i}$ and $M^{p}_{i}$:
\begin{equation}
\label{prototype1}
D^{p}_{i} = Top k \sum^{k} \sum_{m = 0}^{dim}|x^{0}_{i}[m] - x^{0}_{j}[m]|
\end{equation}

\begin{equation}
\label{prototype2}
M^{p}_{i} = |x^{0}_{i}[m] - \frac{1}{k} Top k \sum^{k}_{m = 0} x^{0}_{j}[m]|
\end{equation}
where $dim$ denotes the dimension of $x^0$ vector. We sort $D^{p}_{i}$ and $M^{p}_{i}$, and set the threshold $T = T_r \cdot n_u/ n_p$, where $n_u$ and $n_p$ denote the number of unlabeled and positive samples, respectively. $T_r$ is an adjustable hyper-parameter. 
The smallest $T$ samples of $D^{p}_{i}$ and $M^{p}_{i}$ will be selected in the unlabeled samples
% list 
and only those that satisfy the conditions of both $D^{p}_{i}$ and $M^{p}_{i}$ can be considered as HN samples $\mathcal{I}^{HN}$.

%We treat unlabeled samples with distances higher than the threshold $T$ as Reliable Negative (RN) samples $\mathcal{I}^{RN}$.

\subsubsection{Progressive Fine-tuning}
As vulnerability (positive) samples are often fewer in number compared to samples without vulnerabilities, it is
% it's 
crucial for \tool to continually learn vulnerability patterns from labeled (including pseudo-labeled) samples to obtain more pseudo labels.
We propose progressive fine-tuning to utilize all available unlabeled data and identify more high-quality labeled samples.
%, which consists of two steps: initialization and iteration. 
First, we fine-tune the pre-trained model using labeled samples $\mathcal{I}^{P}$ as positive and HN samples $\mathcal{I}^{HN}$ as negative. The model weights are kept as $\Omega $ and no new samples are added. Then, \tool retains the previous model weights $\Omega $ and determines if any unlabeled samples can be confidently predicted. If so, these samples are labeled as High-quality Positive (HP) $\mathcal{I}^{HP}$ and HN samples $\mathcal{I}^{HN}$ according to the model predicted. Both the HP and HN samples are
% and 
then used as training data for the next epoch.
%\yun{[???]}. 
% These samples are  
This process, called progressive, is repeated by $\mathcal{L}_{ce}$ until the accuracy of validation set
% accuracy 
no longer increases. In the progressive step, the quality of added samples decreases with each training epoch. We reduce the high-quality sample's weights $W^{e}_{i}$ as the number of progressive steps increases,
% as 
calculated as below:

\begin{equation}
\label{Progressive}
W^{e}_{i} = 1 - \frac{e}{E_{m} + 1}, e =\left\{ 1,2,..,E_{m}\right\}
\end{equation}
\begin{equation}
\label{ce}
\mathcal{L}_{ce} = - \sum_{i=1}^{n_e}W^{e}_{i}\hat{y}_{i}log(\hat{p_{i}}), \hat{y}_{i} =  \mathcal{I}^{HN} \cup \mathcal{I}^{HP} \cup \mathcal{I}^{P}
\end{equation}
where $e$ and $E_{m}$ denote the current epoch and progressive fine-tuning training epoch, respectively. $\hat{y}_{i}$ is the concatenation set of the pseudo labels we extract and the positive labels, $\hat{p_{i}}$ denotes the output predicted by the model, and $n_e$ denotes the number of training samples
% train number \yun{[??]} 
in Epoch $e$.

\begin{algorithm}[!tpb]
    \SetAlgoLined
    \footnotesize
    \SetKwInOut{Input}{Input}
    \SetKwInOut{Output}{Output}
    \SetKwInOut{Ensure}{Ensure}
    \SetKwFunction{Inter-class Distance Prototype}{Inter-class Distance Prototype}
    \SetKwProg{Fn}{Function}{:}{}
    \Input{Source Code: $\mathcal{X}$, Selected Label: $\mathcal{C}$, Threshold: $T_{max}, T_{min}$}
    \Output{Pseudo-positive Label: {$\mathcal{I}^{HN}$}, Pseudo-negative Label: {$\mathcal{I}^{HP}$}}
    \Ensure{$\Theta$ denotes the pre-trained model, $\Omega$ denotes the current model}
    \Fn{Distance-aware Label Selection Module}{
    % \tcp{{Start Training: Initial Fine-tuning}} 
    % \For{each training epoch}
    % {
    % \For {each $<x_{i}, c_{i}> \in <\mathcal{X}, \mathcal{C}>$}
    %     {
    %     $<x_{i}^{0}, {h_{i}}, c_{i}> \leftarrow$ Fine Tuning
    %     }
    % }
    
    \tcp{{Start Training:Inter-class Distance Prototype}}  
    
     \For{each $<x_{i}^{0}, c_{i}>$ and $c_{i} = 0$}{
    $<x_{i}^{0}, {h_{i}}, c_{i}> \leftarrow$ Fine Tuning \\
     Initialize $D_{i}$
      \For{each $<x_{j}^{0}, c_{j}>$ and $c_{j} = 1$}
      {
        {Calculated $D^{p}_{i}$ by Eq~\ref{prototype1} given $x_{i}^{0}$ and $x_{j}^{0}$ ; \\}
        {Calculated $M^{p}_{i}$ by Eq~\ref{prototype2} given $x_{i}^{0}$ and $x_{j}^{0}$;}
      }
    } 
    {Sorted $D^{p}_{i}$, $M^{p}_{i}$ and chose Top $K$ samples index as negative samples $\mathcal{I}_{D}^{HN}$, $\mathcal{I}_{M}^{HN}$;\\}
    {Calculated $\mathcal{I}_{D}^{HN} \cap \mathcal{I}_{M}^{HN}$; and chose index as negative samples $\mathcal{I}^{HN}$\\}
    \tcp{Start Training: Progressive Fine-tuning} 
    \For{each training epoch}
    {
    \For{each $<x_{i}, c_{i}> \in <\mathcal{X}, \mathcal{C}>$}
        {
            \If{{$c_{i} = 1$ or $i \in \mathcal{I}^{HN}$}}{$\Omega \leftarrow $ Fine Tuning}
         
        }
    }

    %\tcp{Start Training: Progressive Fine-tuning} 
     \For{each training epoch $e \in \mathcal{E}$}
    {
        \For{each $<x_{i}, c_{i}> \in <\mathcal{X}, \mathcal{C}>$}
        {
            \If{{$c_{i} = 1$ or $i \in \mathcal{I}^{HN}$}}{$\Omega \leftarrow $ Fine Tuning}
         
        }
        \For{each $<x_{i}, c_{i}> \in <\mathcal{X}, \mathcal{C}>$}
        {
            \If{
                {$c_{i} = 0$ and $i \notin \mathcal{I}^{HN}$}
            }
            {
                {$p_{i} \leftarrow $ Prediction;\\} 
                \If{
                    {$p > T_{max} $}
                }
                {\tcp{Add high-quality positive samples;}$i \in \mathcal{I}^{HP}$, $e_{i} \in \mathcal{E}$ ; }

                \If{
                    {$p < T_{min} $}
                }
                {\tcp{Add high-quality negative samples;}$i \in \mathcal{I}^{HN} $, $e_{i} \in \mathcal{E}$ }
            }

        }
    }

    }
    
    \Return {$\mathcal{I}^{HN}$, $\mathcal{I}^{HP}$}

\caption{Inter-class Distance Prototype}
\label{algorithm1}
\end{algorithm}

%\subsubsection{The Algorithm of Affinity from Feature Distance Module}
\subsubsection{The Algorithm of Distance-aware Label Selection Module}

The overall distance-aware label selection module process is depicted in Algorithm \ref{algorithm1}, in which inter-class distance prototype selection and progressive fine-tuning correspond to Lines 2-10 and Lines 11-35, respectively.
The algorithm takes all the training data (including labeled or not) as the input, and outputs
% and outputs are 
pseudo labels (including $\mathcal{I}^{HN}$ and $\mathcal{I}^{HP}$).
% and pre-trained model. 

For inter-class distance prototype selection, the algorithm initializes the representation for each sample (Lines 3). Then, the algorithm selects the inter-class prototype for each unlabeled sample, calculates the corresponding distance, and chooses the HN samples based on the inter-class distance.
% the following operations are performed (Lines 4-6): select the inter-class prototype and calculate the corresponding distance. And then it chooses the HN samples based on the inter-class distance (Lines 9-10) \yun{[???]}.
 For progressive fine-tuning, the algorithm performs fine-tuning again in labeled and HN samples (Lines 19-23), and then selects the HP and HN samples in the training epoch (Lines 24-34).

\subsection{Mixed-supervision representation learning module}
\label{metric_section}
% 选择的样本不具有代表性（数量相对较少），在已有的样本中进一步的挖掘 信息
% 而不是使用有噪声的样本来提升性能
% In our proposed approach we have further improved the performance of \tool by enhancing the inter-class distance and reducing the intra-class. After a sufficient number of iterations, the remaining samples are usually difficult to make the validation set perform better. These samples are more likely to be noisy and cause bias in model learning. 
% To address this problem, we attempt to further learn with the help of pseudo-labels and use the class signature vectors, which represent individual classes in a discriminative manner, to improve the performance of vulnerability detection. 

% In the distance-aware label selection module, \tool selects as many reliable samples as possible and classifies them with pseudo-labels. 
% After a sufficient number of iterations, the remaining samples are more likely to be noisy and lead to biases in model learning. 

%Limited by the fact that pseudo-labels are not fully reliable, 
%

The mixed-supervision representation learning module aims to mitigate the problem of poor-quality labels. %to improve the performance of vulnerability detection.  
% We use an unsupervised loss to mine the relation among different samples of representation and reduce the impact of noisy labeling (pseudo-labeling), \wxc{which enhances the discrimination of the vulnerability representations.} \yun{[how this reflect in the method design?]}. We simultaneously use the weakly supervised loss to construct the relation between representation and label, which enhances the inter-class distance and reduces the intra-class distance for enhancing code representations.
% to enhance the discriminative power of the vulnerability representation. 
We first combine the CE loss and weakly supervised loss to construct the relation between representation and labels, which enhances the inter-class distance and reduces the intra-class distance for enhancing code representations. To alleviate the problem of label noise, we then involve an unsupervised loss to learn the unsupervised representations for reducing the impact of noisy labels.
Specifically, we train $\mathcal{L}_{Metric}$ by minimizing the loss function calculated as below:
% which can be summarized as:

% \begin{equation}
% \label{Intra}
% \mathcal{L}^{Intra}_{i} = \left(\sum_{j=1}^{B}\hat{y}_i ||x^{e}_{i} - x^{e}_{j}|| + (1- \hat{y}_{i}) max\left (m - ||x^{e}_{i} - x^{e}_{j}|| ), 0\right)\right)
% \end{equation}

% \begin{equation}
% \label{Inter}
% \mathcal{L}^{Inter}_{i} = - log \frac{exp\left( z_{i} \cdot z_{\hat{y}_{i}} /\tau \right )}{\sum_{k=1}^{2N}exp\left( z_{i} \cdot z_{k }/\tau \right )}
% \end{equation}

\begin{equation}
\label{supcon}
\mathcal{L}_{Metric} = \alpha \mathcal{L}^{Self}_{i} + (1- \alpha) \mathcal{L}^{Weakly}_{i} + \mathcal{L}_{ce}
\end{equation}
where $\alpha$ is a trade-off parameter. $\mathcal{L}_{ce}$ is calculated as Eq.~\ref{ce}. $\mathcal{L}^{Self}_{i}$ and $\mathcal{L}^{Weakly}_{i}$ \iffalse denotes \fi
denote the relationships in self-supervised and weakly-supervised scenario, respectively. The followings are the details of each loss function.

 Specifically, $\mathcal{L}^{Self}_{i}$ is mainly used to mine the relation of representation itself, 
 % and ignore pseudo-labels
and for each sample $i$, we give a query representation $q$ and a set $\mathcal{B} = \{x_1, . . . x_B \}$ of $B$ samples containing one positive sample and $B - 1$ other samples from the distribution, computed as below:
% the calculated formula is following: 
 
\begin{equation}
\label{cons}
\mathcal{L}^{Self}_{i} = - log \frac{exp\left( q \cdot x_{i} /\tau \right )}{\sum_{k=0}^{B}exp\left( q \cdot x_{k} /\tau \right )}
\end{equation}
where $\tau$ is a temperature hyper-parameter~\cite{DBLP:conf/cvpr/He0WXG20/contra}. The sum is over one chosen sample and $B-1$ contrastive samples. The purpose of the $\mathcal{L}^{Self}_{i}$ is to characterize tries to classify sample $q$ as a chosen sample through a vulnerability sample.

 The weakly supervised loss function $\mathcal{L}^{Weak}_{i}$ is used to further establish a relation between representations and pseudo-labels:

\begin{equation}
\label{supcon}
\mathcal{L}^{Weak}_{i} = - log \frac{exp\left( x_{i} \cdot x_{\hat{y}_{i}} /\tau \right )}{\sum_{k=0}^{B}exp\left( x_{i} \cdot x_{k }/\tau \right )}
\end{equation}
where $\tau$ is also the same hyper-parameter in Eq.~\ref{cons}, and $\hat{y_i}$ denotes the pseudo-label of sample $i$.
% And the $\mathcal{L}_{ce}$ is calculated as Eq.~\ref{ce}.

%% file: Sections/4_Evaluation.tex
In this section, We evaluate the \tool %with the state-of-the-art vulnerability methods 
and aim to answer the following research questions (RQs):

% \begin{enumerate}[label=\bfseries RQ\arabic*:,leftmargin=.5in]
%     \item How does \tool perform in vulnerability detection compared with the positive and unlabeled learning approaches?
%     \item How does \tool perform in vulnerability detection compared with previous approaches?
%     \item What is the influence of different modules on the detection performance of \tool?
%     \item What is the impact of hyper-parameters on the performance of \tool?
% \end{enumerate}

% \dw{I acutally recommend to swtich the current RQ1 and RQ2 as follows: 
% The rationale is that we first prove that detection with PU learning can outperform the SOTAs in vul detection, then study the performance of the different PU strategies, next study the impact of different components}
% \dw{here, the order of RQs is not consistent with the following content.}
\begin{enumerate}[label=\bfseries RQ\arabic*:,leftmargin=.5in]
    
    \item How does \tool perform in vulnerability detection with different weakly supervised methods in PU settings?
    \item How does \tool perform %in vulnerability detection (VD) 
    compared with the state-of-the-art vulnerability detection approaches?
    \item What is the influence of different modules on the detection performance of \tool?
    \item How do the different hyper-parameters impact the performance of \tool?
    %What is the impact of hyper-parameters on the performance of \tool?
\end{enumerate}

%\subsection{Experiment Setup}
\subsection{Datasets}
% Our experimental study is based on three widely used code vulnerability datasets, including FFMPeg+Qemu~\cite{devign}, Reveal~\cite{reveal} and Fan et al.s~\cite{DBLP:conf/msr/FanL0N20/fan}. The manually annotated dataset FFMPeg + Qemu used in the Devign study is built on four different open source C projects with a total count of 22k+ and about 45\% of the data entries are vulnerable. The dataset provided by Fan et al.'s is one of the largest vulnerability datasets, which is based on a large number of open source github projects with a total count of 188k+ and about 5.7\% of the vulnerability entries. The Reveal dataset contains a range of open source software applications with a total of 18k+ functions and about 9.2\% of the functions are flawed. 
% {\dw{pls add a table to show the basic stats of the above datasets. also please add refs for datasets}
To answer the questions above, we %our experimental study 
choose three widely-used vulnerability datasets, including FFMPeg+Qemu~\cite{devign}, Reveal~\cite{reveal} and Fan et al.~\cite{DBLP:conf/msr/FanL0N20/fan}. 
FFMPeg+Qemu consists of two open-source C projects with a total of 22k samples, out of which 10k samples are vulnerable.
The Reveal dataset tracks historical vulnerabilities in two open-source projects, with over 22k and approximately 2k vulnerable samples.
Fan \et's dataset collects 91 types of vulnerabilities from 348 open-source GitHub projects, with around 188k total samples and 10k vulnerable samples.
%The FFMPeg+Qemu is built on two different open-source C projects with a total count of 22k, of which about 10k samples are vulnerable. The Reveal dataset is created by tracking historical vulnerabilities in two open-source projects, with a total of over 22k entries and around 2k vulnerable entries.  The dataset provided by Fan \et collected 91 different types of vulnerabilities from 348 open source Github projects. This dataset contains approximately 188k total entries and 10k vulnerable entries. 

\subsection{Baselines}
In this paper, %the software vulnerability detection task, 
we compare \tool with four representative weakly supervised learning methods in the PU setting and five state-of-the-art vulnerability detection methods in the supervised setting.
%We compare PUVD with typical vulnerability detection methods as well as Positive and Unlabeled learning methods. 

In RQ1, we compare \tool with four weakly supervised learning methods in the PU setting: 
%\sz{[why choose these four methods? popular? sota? representative?]}
%\wangxc{[CR-SVM is the classical method in the Two Step Techniques, uPU and nnPU are representative methods for unbiased PU learning, and self-PU reaches SOTA in this field.]}
\begin{enumerate}
%1) 
\item \textbf{Cosine and Rocchio SVM} (CR-SVM)~\cite{DBLP:conf/emnlp/LiLN10}: CR-SVM extracts tf-idf and uses the cosine similarity between unlabeled and labeled samples. Then it uses the iterative SVM to select the optimal classifier. 

%2) 
\item \textbf{Unbiased PU} (uPU)~\cite{DBLP:conf/nips/PlessisNS14}: 
%The unbiased PU 
uPU treats the unlabeled sample as the sum of positive and negative samples with different weights, and constructs a risk estimator to train.
% An unbiased learning method based on unbiased risk estimator. 

%3) 
\item \textbf{Non-negative PU} (nnPU)~\cite{DBLP:conf/nips/KiryoNPS17}:
nnPU constructs a non-negative risk estimator based on the uPU, which adds a limitation to the loss function of risk estimator to alleviate the overfitting problems.

% A large-scale PU algorithm based on a non-negative risk estimator, which can effectively solve the negative empirical risk and overfitting problems in Unbiased PU.

%4) 
\item \textbf{Self-PU}~\cite{DBLP:conf/icml/ChenCCYG0W20}: 
Self-PU is 
%\sz{is the state-of-the-art PU learning method that employs} %proposes 
a self-paced learning algorithm, self-calibrated instance-awarded loss, and self-distillation strategy to train the model.
%This approach combines self-training with PU learning and builds three different functional "self" oriented modules into the self-training.
\end{enumerate}

%\dw{In RQ1, we compare \tool with four VD SOTAs:}
%In RQ2, we compare \tool with four VD SOTAs:
In RQ2, we adopt five state-of-the-art vulnerability detection approaches for comparison in the supervised learning setting, including:
\begin{enumerate}
%1) 

\item \textbf{SySeVR}~\cite{sysevr}: SySeVR uses statements, program dependencies, and program slicing generated from source code, and utilizes a bidirectional recurrent neural network to vulnerability detection.

\item \textbf{Devign}~\cite{devign}: Devign constructs a joint graph by Abstract Syntax Tree (AST), CFG, DFG and Natural Code Sequence (NCS) and uses GGNN for vulnerability detection.

%2) 
\item \textbf{Reveal}~\cite{reveal}: Reveal divides vulnerability detection into two steps: feature extraction steps by GGNN and training steps by multi-layer perceptron and triplet loss.

%3) 
\item \textbf{IVDetect}~\cite{IVDETECT}: IVDetect constructs a program dependency graph and utilizes a feature-attention graph convolutional network to learn the graph representation.

%4) 
\item \textbf{LineVul}~\cite{DBLP:conf/emnlp/FengGTDFGS0LJZ20/codebert}: LineVul trains on the Transformer architecture. To ensure fairness, we use the word-level tokenizer version for comparison.
\end{enumerate}

%Specifically, in the case of full marking, the vulnerability detection methods for comparison are: 

\subsection{Implementation Details}
To ensure the fairness of the experiment, we use the same data split for all approaches.
In the PU %positive and unlabeled 
learning scenario, we randomly label %labeled 
30\% of the training positive samples based on the assumptions (Section~\ref{sec:background}).
We repeat the labeling scenario three times and use the mean as the experimental results.
In the supervised learning scenario, we randomly partition the datasets into disjoint training, validation, and test sets in a ratio of 8:1:1. 

We try our best to reproduce all baseline models from publicly available source code and papers, and use %used 
the same hyper-parameter settings as in the original text whenever possible.

We fine-tune the pre-trained model CodeBERT~\cite{DBLP:conf/emnlp/FengGTDFGS0LJZ20/codebert} with a learning rate of $2e-5$. The batch size is set to $32$. The top nearest samples $k$ is set to 30\% to choose prototype samples. 
% (\dw{so, 30\% of labeled samples are used as the top nearest samples?} \wxc{A: yes}) 
The $T_r$ is set to 0.3. During each fine-tuning, we train our model on a server with NVIDIA A100-SXM4-40GB for maximum epochs $E_m$ of 5. 
% (\dw{what the ratio is? } \wxc{A: Ratio \ie $T_r$ an adjustable hyper-parameter set under the Eq.7})

\subsection{Evaluation Metrics}
We use the following four commonly-used metrics to measure \tool's performance:

\textbf{Precision:} $Precision = \frac{TP}{TP+FP}$. The precision measures the percentage of true vulnerabilities out of all the vulnerabilities that are retrieved. $TP$ and $FP$ denote the number of true positives and false positives, respectively.
% \begin{equation}
% \label{precision}
% Precision = \frac{TP}{TP+FP}
% \end{equation}

\textbf{Recall:} $Recall = \frac{TP}{TP+FN}$. The recall measures the percentage of vulnerable sample that are retrieved out of all vulnerable samples. $TP$ and $FN$ denote the number of true positives and false negatives, respectively.

% \begin{equation}
% \label{recall}
% Recall = \frac{TP}{TP+FN}
% \end{equation}

\textbf{F1 score:} $F1\ score = 2 \times \frac{Precision\times  Recall}{Precision+Recall}$. The F1 score is the harmonic mean of precision and recall metrics.

% \begin{equation}
% \label{f1}
% F1\ score = 2 \times \frac{Precision\times  Recall}{Precision+Recall}
% \end{equation}

\textbf{Accuracy:} $Accuracy = \frac{TP+TN}{TP+TN+FN+FP}$. The accuracy measures the percentage of correctly classified samples out of all samples.
$TN$ represents the number of true negatives and $TP+TN+FN+FP$ represents the number of all samples.

% \begin{equation}
% \label{Acc}
% Accuracy = \frac{TP+TN}{TP+TN+FN+FP}
% \end{equation}

%% file: Sections/5_Experimental_Result.tex
\input{Tables/RQ1}

\subsection{RQ1. {Effectiveness of \tool in PU setting}}
To answer RQ1, we compare \tool{} with the four PU learning baseline methods with the four performance metrics (\ie accuracy, precision, recall, and F1 score) on the three datasets. 
% In each dataset, 30\% of the positive samples are randomly selected as the labeled samples for the positive and unlabeled task settings \sz{[this has been mentioned in the last section?]}. 
To ensure fairness in the experiment, the same labels are chosen for all the baseline methods, and we choose the labeled samples 3 times %(\dw{how many}) 
and report the averaged results. %take the mean as the reported results. 
Table~\ref{rq1} shows the results, and the performance of \tool is shown in the bottom row.

%Our study finds 
From Table~\ref{rq1}, we can find that \tool{} outperforms 
%performs better than 
all PU learning methods in detecting software vulnerabilities across all three datasets and obtains the best results in 11 out of 12 metrics. %as measured by Recall and F1 score metrics across all three datasets. %In fact, 
%When considering all 12 combined cases, \tool{} outperforms other baselines in 10 \sz{11?} cases and ranks within the top three in all cases. 
Specifically, \tool{} achieves absolute improvements of 2.78\%, 18.44\%, and 18.93\%  over the F1 scores of the best baseline method on the FFMPeg+Qemu~\cite{devign}, Reveal~\cite{reveal} and Fan \et~\cite{DBLP:conf/msr/FanL0N20/fan} datasets, respectively. As for accuracy metric, \tool{} outperforms baseline methods by at least 2.31\%, 0.89\%, and 0.47\% on these three datasets, respectively.
%with corresponding relative improvements of 2.31\%, 0.89\% and 10.41\% in accuracy. 
In the unbalanced dataset (\ie Reveal and Fan \et), \tool achieves better performance in all four metrics (eight situations). This is due to the ability of \tool to better extract high-quality pseudo-labels from unbalanced samples and detect vulnerabilities. 
% Although CR-SVM has achieved better performance than \tool in the recall metric in the FFMPeg+Qemu dataset, compared with it, \tool improves the performance by 9.39\%, 8.83\% and 2.78\% respectively in other three metrics.
Overall, \tool can be able to detect more vulnerable samples and therefore proved to be more effective in PU settings.

% % \sz{[change here, inconsistent with table]} 
% In the FFMPeg+Qemu dataset, CR-SVM has achieved better performance than \tool in the recall metric. 
% The reason is that the FFMPeg+Qemu dataset has a balanced situation, with the number of non-vulnerable samples being similar to the number of vulnerable samples, which leads to the di
% In contrast, \tool can be able to detect more vulnerable samples and therefore proved to be more effective overall. 
% It indicates that \tool can identify more vulnerabilities than weakly supervised methods in positive and unlabeled setting scenarios. 

%It indicates that \tool can identify more vulnerabilities than PU learning methods in positive and unlabeled task scenarios. 

% (\dw{here, it is a little bit confusing. To me, the results should support the argument: the proposed PUVD compared with other PU learning approaches is more effective in detecting vulnerability. The current description sounds like the PUVD is better than SOTAs in vul detection, which is RQ2 actually})

Our results also show that the existing weakly supervised learning methods have much potential for improvement in the area of software vulnerability detection. They ignore the quality of the extracted pseudo-labels and the possible presence of noise in the vulnerability labels. Overall, our results demonstrate the superior performance of \tool{} in detecting vulnerabilities, highlighting its potential to improve the effectiveness of computer security measures.

% \dw{can you also add some results to show the accuracy of label learning of 4 different PU approaches? not just on vulnerability detection to show that the proposed PU can learn higher quality labels than others}

 \begin{tcolorbox}
 \textbf{Answer to RQ1:} In the PU setting, \tool outperforms all the baseline methods in terms of precision and F1 score. 
 In particular, \tool{} achieves 2.78\%, 18.44\%, and 18.93\% improvements in F1 score over the best-performing baseline method on the three datasets, respectively. 
 \end{tcolorbox}

\input{Tables/RQ2}

\subsection{RQ2. {Effectiveness of \tool in supervised setting}}
% \dw{why not add your newest approach AMPLE into the analysis? also if you have time, it is good to add LineVD and LineVul}

To answer RQ2, we %then 
compare \tool{} with the five vulnerability detection methods on the three datasets in supervised learning settings. 
All baselines use the labels of the positive and negative samples to train the binary classifier. 

Table~\ref{rq2} shows the results of the vulnerability detection baselines. %Specifically, 
We can find that \tool{} has the best performance in 8 out of 12 cases. 
For example, on Reveal, \tool outperforms the best baseline methods by 1.45\%, 5.51\%, 21.24\%, and 12.46\% regarding the accuracy, precision, recall, and F1 score, respectively.
%and achieves the best results on four metrics in the Reveal dataset, which improves the performance by 3.37\%, 13.56\%, 31.07\%, and 20.02\% respectively compared to the best-performing methods. 
Compared with %to 
the average of previous vulnerability detection methods, \tool achieves %has 
absolute improvement of  %absolutely improved by 
4.61\%, 7.59\%, 22.32\%, and 14.88\%  with respect to the four metrics, respectively.
% \sz{Both} Reveal and Devign %each has
% have a metric value that beats \tool{}, \sz{[no need to emphasize this?]} but the superiority is on only one dataset, with poor performance on the remaining metrics.
% It may be due to that \tool does not treat all the unlabeled samples as negative samples. With less label noise in the dataset, the model may actually identify incorrect pseudo-labels leading to reduced performance.
%Our results also show that the graph-based model approach has less difference in performance under balanced data sets and CodeBERT, while the results under unbalanced data sets are much worse than CodeBERT. We believe that the structural information (\eg, data flow graph and control flow graph) of source code that is more easily captured by the graph-based model approach under balanced conditions, unbalanced data scenarios easily disrupt the structural information.

Our results also show that the graph-based approaches (Devign, Reveal, and IVDetect) perform similarly to \tool when the samples with vulnerabilities or not are balanced. 
However, in scenarios where the datasets are unbalanced, the performance of graph-based approaches is worse than \tool. It may be due to the fact that the graph-based model approaches are hard to capture structural information  (\eg, data flow graph and control flow graph) in unbalanced scenarios~\cite{9656689}.

The improvement of the experiment is non-trivial, the \tool only selects positive and unlabeled samples in the PU setting. In contrast, all of the vulnerability detection baselines use all labels in the training process.
%captures structural information  (\eg, data flow graph and control flow graph) harder in unbalanced scenarios.
% , but unbalanced data scenarios can disrupt this structural information.

 \begin{tcolorbox}
 \textbf{Answer to RQ2:} In the vulnerability detection, 
 \tool performs better in most cases. Compared to existing methods, the \tool{} performs 4.61\%, 7.59\%, 22.32\%, and 14.88\% improvements on average on the four metrics, respectively. 
 \end{tcolorbox}

\subsection{RQ3. {Effectiveness of different components in \tool}}

%To answer RQ3, 
In this section, we explore the impact of different components on the performance of \tool. {Specifically, we study the two involved modules including the Distance-aware Label Selection (DLS) module and Mixed-supervision Representation Learning
module (MRL) module.}
\input{Tables/RQ3}

\subsubsection{Distance-aware Label Selection Module}
To explore the contribution of the DLS module, we create two variants of \tool without inter-class distance prototype (\ie ID Prototype) and %or 
progressive fine-tuning (\ie PFine-tuning), respectively. The other settings of these two vairants
are consistent with \tool. The purpose of the DLS module is to select high-quality labels from unlabeled samples, which requires at least one of the components to construct a binary classifier. Therefore, we separate two variants for the ablation experiments.

As shown in Table~\ref{table_rq3}, both ID Prototype and PFine-tuning components can improve the performance of \tool on all datasets. 
Specifically, on the unbalanced datasets (\ie Reveal and Fan \et), PFine-tuning component achieves an average improvement of 27.76\% and 16.07\% in terms of accuracy and F1 score, respectively; while ID Prototype only boosts by 13.27\% and 9.89\%. This indicates that PFine-tuning has a greater effect on the unbalanced dataset.
%The results show that PFine-tuning has a greater effect on the unbalanced dataset (\ie Reveal and Fan \et), which has improved by 27.76\% and 16.07\% on average in terms of accuracy and F1 score, respectively, while ID Prototype has only 13.27\% and 9.89\% boosts. 
Conversely, on the balanced dataset (\ie FFMPeg+Qemu), ID Prototype obtains improvements of 3.00\% and 9.88\% respectively, which outweighs the improvements of PFine-tuning. 
%\wxc{(i.e. xxx)}. 
The results demonstrate that different components of the DLS module focus on the different situations to select high-quality samples, 
% can promote the construction of a high-quality binary classifier and 
which benefit the performance of vulnerability detection.%\sz{[this sentence is confusing]}

\subsubsection{Mixed-supervision Representation Learning
Module}

To understand the impact of MRL module, we also deploy a variant of \tool without the MRL module. 
%Due to 
Since the above module already constitutes the training of classifier,
%a classifier training, 
the variant (\ie w/o MRL) directly eliminates the training of the MRL module.

Table~\ref{table_rq3} shows the performance of the variant on the three datasets.
Overall, the accuracy, and F1 score in three datasets achieve higher values with the addition of the MRL module.
The MRL module improves the accuracy by 0.59\%, 5.36\%, and 9.85\% on FFMPeg+Qemu, Reveal, and Fan \et. datasets, respectively. As for the F1 score, the MRL module also brings an improvement of 1.74\%, 5.98\%, and 4.51\% on these three datasets, respectively.
%The accuracy of MRL module improves performance by 0.59\%, 5.36\%, and 9.85\% on FFMPeg+Qemu, Reveal, and Fan \et. datasets, respectively. \tool also increases the F1 score by 1.74\%, 5.98\%, and 4.51\%, respectively. 
%Compared with the variant without the MRL module, 
The results indicate that MRL module can enhance the discriminative power of the vulnerability representation and bring a performance improvement in vulnerability detection. %It brings a performance improvement in software vulnerability detection.

\begin{tcolorbox}
\textbf{Answer to RQ3:} Both DLS and MRL modules contribute significantly to the performance of \tool.
The DLS module average boost the F1 score performance of 4.47\%, 17.22\%, and 8.74\% on the three datasets, respectively.
The MRL module improves \tool{} by 1.74\%, 5.98\%, and 4.51\%, respectively.
%is improved by 1.74\%, 5.98\%, and 4.51\%, respectively. 
\end{tcolorbox}

\subsection{RQ4: Influences of Hyper-parameters on \tool{}}

To answer RQ4, we explore the impact of different hyper-parameters, including the labeling ratio and the top $k$ samples chosen in the inter-class distance prototype.

\subsubsection{Ratio of positive labeling}

Table~\ref{rq4_ratio} shows the performance of %four metrics of 
\tool on four metrics with different ratios of positive labeling samples. It means that all negative samples and unlabeled positive samples are considered unlabeled samples. 
As the ratio of labeling increases, the performance of \tool also increases gradually.
\tool achieves the best performance when all the positive samples are used as labeling samples.
Compared with the performance of the 10\% labeling ratio, the 100\% labeling ratio can improve it by %the performance of 
14.13\%, 20.62\%, 11.00\%, and 20.80\% on four metrics, respectively.
%It means 
It demonstrates that it is important to have a larger number of positive samples to improve the discriminative ability of prototypes formed by positive samples. Additionally, extracted high-quality negative samples are more believable by increasing positive samples, leading to better overall performance.
 
When dealing with unbalanced datasets such as Reveal and Fan, we observe that performance increases faster when the proportion of labeled samples is below 30\%. However, the growth rate of performance decreases as the number of labeled samples continues to increase.
In contrast, the growth of performance in FFMPeg+Qemu is more balanced since the limited number of negative samples makes it more difficult to capture high-quality labeled samples.

\input{Tables/RQ4_ratio}

\subsubsection{Top $k$ samples of inter-class distance prototype} 
We also explore the effect of the number of Top $k$ samples of the inter-class distance prototype of \tool.
As shown in Table~\ref{rq4_k}, it shows the number and its corresponding accuracy in selecting high-quality negative samples. The higher the accuracy of the recognition, the higher quality of the identified labeled samples is.% the high-quality of the identified confidence-able samples.

Overall, the larger the number of $k$ will lead to higher accuracy. However, the number of labels with $k$ greater than 30\% ratio of labeled samples reaches the recognition accuracy of 75.18\%, 99.34\%, and 99.24\% in the three datasets, respectively, and the growth rate gradually converges to 0. Therefore, in order to reduce computational resource consumption, we choose the 30\% ratio of labeled samples as the value of $k$.

In addition, we also present the number of high-quality negative samples (\ie Num) in Table V.
%Num denotes the number of selecting high-quality negative samples in the inter-class distance prototype. \
From the results, we can observe that different selections of $k$ lead to different sum and mean of the prototype distance (\ie $D^{p}_{i}$ and $M^{p}_{i}$) for sample $i$, leading to differences in the number of high-quality samples selected. For example, the $k$ value of 100\% selects 7.18\%, 7.74\%, and 11.76\% more high-quality samples than the value of 3.
%It means that different selections of $k$ lead to the sum and mean of the prototype distance (\ie $D^{p}_{i}$ and $M^{p}_{i}$) for sample $i$ differ, leading to differences in the number of high-quality samples selected. 
Overall, the higher the value of $k$, the smaller the difference between the $D^{p}_{i}$ and $M^{p}_{i}$ is, and the more high-quality samples are selected.
\input{Tables/RQ4_k}

 \begin{tcolorbox}
 \textbf{Answer to RQ4:} 
In vulnerability detection, the \tool's performance can be affected %effected 
by the ratio of labeling and top samples. However, our default settings have been optimized for 
% optimal 
results.
\end{tcolorbox}

%% file: Tables/RQ1.tex
\begin{table*}[h]
\centering

\setlength{\tabcolsep}{1.2mm}
\renewcommand{\arraystretch}{1.2}

\caption{
Comparison results between \tool and the weakly supervised learning approaches in the PU setting on the three datasets. %The best result for each metric is highlighted in bold. 
The shaded cells represent the performance of the best methods in each metric. Dark cells represent the best performance.}
\resizebox{.97\textwidth}{!}{
\begin{tabular}{l|cccc|cccc|cccc}
\toprule
\diagbox{Metrics(\%) }{Dataset} & \multicolumn{4}{c|}{FFMPeg+Qemu ~\cite{devign}}        & \multicolumn{4}{c|}{Reveal~\cite{reveal}}             & \multicolumn{4}{c}{Fan \et~\cite{DBLP:conf/msr/FanL0N20/fan}}                \\
\midrule
Baseline                         & Accuracy & Precision & Recall & F1 score     & Accuracy & Precision & Recall & F1 score    & Accuracy & Precision & Recall & F1 score    \\
\midrule
% Cosine and Rocchio SVM~\cite{DBLP:conf/emnlp/LiLN10}                    & 49.19                & 45.56                & 54.34                & 49.56                & 85.58                & 13.16                & 6.15                 & 8.38                 & 85.34                & 6.06                 & 11.18                & 7.86                 \\
% Unbiased PU~\cite{DBLP:conf/nips/PlessisNS14}                 & 55.17                & 51.54                & 41.35                & 45.74                & 78.66                & 21.43                & 32.70                & 24.71                & 89.74                & 14.69                & 18.38                & 16.26                \\
% Non-negative PU~\cite{DBLP:conf/nips/KiryoNPS17}                  & 55.59                & 52.92                & 30.41                & 38.57                & 78.66                & 21.43                & 32.70                & 24.71                & 89.74                & 14.69                & 18.38                & 16.26                \\
% Self-PU~\cite{DBLP:conf/icml/ChenCCYG0W20}                & 57.72                & 54.53                & 47.97                & 51.04                & 84.92                & 28.76                & 27.46                & 28.09                & 83.09                & 10.10                & 25.59                & 14.48                \\          

Cosine and Rocchio SVM~\cite{DBLP:conf/emnlp/LiLN10}                     & 48.99                   & 45.83                   & \cellcolor{gray!70}60.80                   & 52.21                  & 86.10                   & 7.87                    & 9.12                    & 7.80                   & 85.66                   & 9.57                    & 6.18                    & 6.72                   \\
Unbiased PU~\cite{DBLP:conf/nips/PlessisNS14}                  & 56.08                   & 54.43                   & 30.15                   & 38.74                  & 83.09                   & 28.79                   & 26.55                   & 24.75                  & 91.44                   & 24.52                   & 11.22                   & 10.62                  \\
Non-negative PU~\cite{DBLP:conf/nips/KiryoNPS17}                  & 56.01                   & 50.20                   & 41.96                   & 46.60                  & 83.07                   & 28.93                   & 27.18                   & 25.38                  & 91.45                   & 25.15                   & 10.77                   & 9.79                   \\
Self-PU~\cite{DBLP:conf/icml/ChenCCYG0W20}               & 53.69                   & 50.15                   & 44.30                   & 45.85                  & 84.41                   & 22.01                   & 17.32                   & 18.61                  & 89.52                   & 19.75                   & 9.39                    & 9.04                   \\

\midrule
\tool & \cellcolor{gray!70}58.38                   & \cellcolor{gray!70}54.66                   & 55.48                   & \cellcolor{gray!70}54.99                  & \cellcolor{gray!70}86.99                   & \cellcolor{gray!70}40.83                   & \cellcolor{gray!70}47.34                   & \cellcolor{gray!70}43.82                  & \cellcolor{gray!70}91.92                   & \cellcolor{gray!70}29.05                   & \cellcolor{gray!70}30.30                   & \cellcolor{gray!70}29.55   \\

\bottomrule

\end{tabular}}

\label{rq1}
\end{table*}

%% file: Tables/RQ2.tex
\begin{table*}[h]
\centering

\setlength{\tabcolsep}{1.2mm}
\renewcommand{\arraystretch}{1.2}

\caption{
Comparison results between \tool and the supervised vulnerability detection methods on the three datasets. ``-'' means that means that the baseline fails to converge in this scenario. The best result for each metric is highlighted in bold. 
%The shaded cells represent the performance of the top-1 best methods in each metric. Darker cells represent better performance.
}
\resizebox{.97\textwidth}{!}{
\begin{tabular}{l|cccc|cccc|cccc}
\toprule
\diagbox{Metrics(\%) }{Dataset} & \multicolumn{4}{c|}{FFMPeg+Qemu \cite{devign}}        & \multicolumn{4}{c|}{Reveal \cite{reveal}}             & \multicolumn{4}{c}{Fan \et \cite{DBLP:conf/msr/FanL0N20/fan}}                \\
\midrule
Baseline                         & Accuracy & Precision & Recall & F1 score     & Accuracy & Precision & Recall & F1 score    & Accuracy & Precision & Recall & F1 score    \\
\midrule
% VulDeePecker                    & 49.61   & 46.05    & 32.55 & 38.14 & 76.37   & 21.13    & 13.10 & 16.17 & 81.19   & \cellcolor{gray!70}\textbf{38.44}    & 12.75 & 19.15 \\

% Russell \et                  & 57.60   & 54.76    & 40.72 & 46.71 & 68.51   & 16.21    & 52.68 & 24.79 & 86.85   & 14.86    & {26.97} & 19.17 \\

SySeVR                          & 47.85   & 46.06    & 58.81 & 51.66 & 74.33   & 40.07  & 24.94 & 30.74   & 90.10   & 30.91   & 14.08 & 19.34 \\

Devign                          & 56.89   & 52.50    &  64.67 &  57.95 & 87.49   & 31.55    & 36.65 & 33.91 & 92.78   & 30.61   & 15.96   & 20.98 \\

Reveal                          & 61.07   &  55.50    &  \cellcolor{gray!70}70.70 & 62.19 &  81.77   &  31.55    & {61.14}& 41.62   & 87.14   & 17.22   & 34.04 & 22.87 \\

IVDetect                        & 57.26   & 52.37    & 57.55 & 54.84 & -   & -         & -      & -      & -   & -         & -      & -      \\

LineVul                      & 62.37                         & \cellcolor{gray!70}61.55                        & 48.21                        & 54.07 & 87.51            & 43.63            & 56.15            & 49.10             & \cellcolor{gray!70}94.44            &  \cellcolor{gray!70}50.5             & 28.53            & 36.46            \\

\midrule
%Our proposed         
\tool                    & \cellcolor{gray!70}63.14                         & 58.23                        & 69.88                        & \cellcolor{gray!70}63.53 & \cellcolor{gray!70}88.96            & \cellcolor{gray!70}49.14            & \cellcolor{gray!70}82.38            & \cellcolor{gray!70}61.56            & 92.70            & 38.00            & \cellcolor{gray!70}42.56            & \cellcolor{gray!70}39.46          
\\
\bottomrule
\end{tabular}}

\label{rq2}
\end{table*}

%% file: Tables/RQ3.tex
\begin{table}[t]
\centering

\setlength{\tabcolsep}{1.4mm}
\renewcommand{\arraystretch}{1.0}
\caption{Impact of the different components on the performance of \tool.}
\resizebox{.45\textwidth}{!}{
\begin{tabular}{c|l|cc}
\toprule
\multicolumn{1}{c|}{Dataset} & Module               & Accuracy             & F1 score                  \\
\midrule
\multirow{4}{*}{FFMPeg+Qemu}     & w/o ID Prototype   & 56.30              & 47.35          \\
                            & w/o PFine-tuning        & 56.77              & 58.17        \\
                            & w/o MRL             & 58.71                & 55.49                \\
                            & \tool & 59.30                & 57.23                \\
\midrule
\multirow{4}{*}{Reveal}     & w/o ID Prototype              & 73.97                & 41.27                \\
                            & w/o PFine-tuning               & 60.47                & 32.66                \\
                            & w/o MRL             & 80.43                & 48.20                \\
                            & \tool & 85.79                & 54.18                \\
\midrule
\multirow{4}{*}{Fan}        & w/o ID Prototype             & 72.90                & 21.83                \\
                            & w/o PFine-tuning                & 57.43                & 18.09                \\
                            & w/o MRL             & 77.77                & 24.19                \\
                            & \tool & 87.62                & 28.70   \\        
\bottomrule
\end{tabular}}

\label{table_rq3}
\end{table}

%% file: Tables/RQ4_ratio.tex
\begin{table*}[h]
\centering

\setlength{\tabcolsep}{1.2mm}
\renewcommand{\arraystretch}{1.2}

\caption{The effect of different ratios of labeled samples on the performance of the \tool.}
\resizebox{.97\textwidth}{!}{
\begin{tabular}{l|cccc|cccc|cccc}
\toprule
{Dataset} & \multicolumn{4}{c|}{FFMPeg+Qemu \cite{devign}}        & \multicolumn{4}{c|}{Reveal \cite{reveal}}             & \multicolumn{4}{c}{Fan \et \cite{DBLP:conf/msr/FanL0N20/fan}}                \\
\midrule
Ratio                         & Accuracy & Precision & Recall & F1 score     & Accuracy & Precision & Recall & F1 score    & Accuracy & Precision & Recall & F1 score    \\
\midrule
10\% & 55.93                         & 51.93                         & 54.66                         & 53.26 & 64.69 & 17.46 & 61.48 & 27.20  & 82.08 & 14.08 & 43.22 & 21.24 \\
30\% & 58.38 & 54.66 & 55.48 & 54.99 & 83.96 & 35.11 & 47.40 & 39.75 & 90.17 & 23.97 & 34.84 & 28.40 \\
50\%                         & 59.77                         & 56.54                         & 53.71                         & 55.09 & 84.43 & 39.46 & 84.43 & 53.79 & 87.20& 22.11 & 51.09 & 30.86 \\
70\%                         & 60.65                         & 57.38                         & 55.78                         & 56.57 & 86.41 & 43.10  & 83.20  & 56.78 & 90.01 & 27.85 & 49.38 & 35.61 \\
100\%                        & 63.14                         & 58.23                         & 69.88                         & 63.53 & 88.96 & 49.14 & 82.38 & 61.56 & 92.99 & 37.97 & 40.10 & 39.00  
\\
\bottomrule
\end{tabular}}

\label{rq4_ratio}
\end{table*}

%% file: Tables/RQ4_k.tex
\begin{table}[h]

\centering

\setlength{\tabcolsep}{1.mm}
\renewcommand{\arraystretch}{1.1}
\caption{The number of samples $k$ selected by \tool and the corresponding accuracy in the inter-class prototype step. The percentage (\%) indicates the number of selected samples among all unlabeled samples. ``Num'' denotes the number of selecting reliable negative samples in the inter-class prototype step.}
\resizebox{.45\textwidth}{!}{
\begin{tabular}{c|cc|cc|cc}
\toprule
{Dataset} & \multicolumn{2}{c|}{Devign} & \multicolumn{2}{c|}{Reveal} & \multicolumn{2}{c}{Fan} \\
\midrule
$k$      & Acc(\%)     & Num     & Acc(\%)      & Num     & Acc(\%)      & Num    \\
\midrule
3      & 73.03         & 5210       & 99.21         & 4822       & 95.67       & 39564     \\
5      & 73.25         & 5211       & 99.20          & 4770       & 95.67       & 39438     \\
30\%   & 75.18         & 5560       & 99.34         & 5172       & 99.24       & 44264     \\
50\%   & 75.25         & 5563       & 99.42         & 5194       & 99.26       & 44279     \\
100\%  & 75.30          & 5584       & 99.44         & 5195       & 99.28       & 44217     \\
\bottomrule
\end{tabular}}
\label{rq4_k}
\end{table}

%% file: Sections/6_Discussion.tex
\subsection{Why does \tool Work?}

\begin{figure*}[ht]
	\centering
	
	\includegraphics[width=0.95\textwidth]{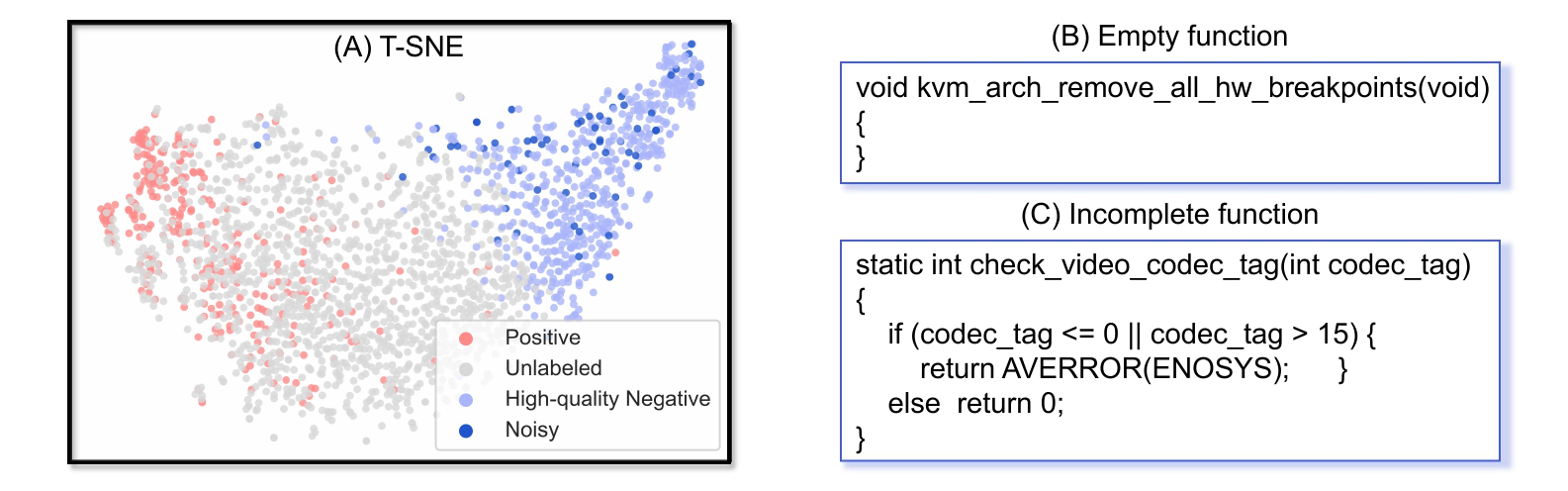}
	\caption{
	(A) illustrates the T-SNE~\cite{van2008visualizing/TSNE} distribution between vulnerable (pink), high-quality non-vulnerable samples (blue), unlabeled (grey) and part of noisy examples (dark blue) in the FFMPeg+Qemu dataset.
	(B) and (C) denote the two different types of incorrectly labeled samples, respectively. }
	\label{disscussion}

% \dw{Figure B and C are too big. the size should be reduced if the space is tight}
 
\end{figure*}

We identify the advantages of \tool, which can explain its effectiveness in software vulnerability detection. We also show the two types of incorrectly labeled samples in FFMPeg+Qemu~\cite{reveal}.

\textbf{(1) \tool is able to identify high-quality samples.}
The proposed DLS module helps \tool to select high-quality samples from unlabeled samples, which involves the inter-class distance prototype and progressive fine-tuning to identify high-quality samples. As shown in Figure~\ref{disscussion}(A), the T-SNE of sample distribution illustrates the distinct class discriminability between positive samples and high-quality negative samples. It demonstrates the effectiveness of \tool.
The experimental results show that \tool can be able to identify 75.18\%, 99.34\%, and 99.24\% high-quality samples in FFMPeg+Qemu, Reveal, and Fan \et, respectively.
% This shows that \tool is able to identify the vast majority of correct unlabeled samples and obtain good performance in software vulnerability detection.
Overall, as the number of labeled samples increases, the number of high-quality samples and the accuracy rate also increase. 

\textbf{(2) \tool well leverages the mislabeled data.}
% In addition, to identify high-quality samples,
% identified reliable samples, 
\tool also pinpoints
% points out 
the labels that are originally mislabeled in the FFMPeg+Qemu~\cite{devign}.
% \tool also mislabeled on the three datasets. 
Specifically, we explore the incorrectly identified high-quality samples in \tool. By manually examining the code and corresponding labels. We randomly select 200 code samples which are associated with different pseudo-labels generated by \tool and ground truth from
% in 
FFMPeg+Qemu for manual checking.
% to conduct the manual inspection. These samples have different \tool and ground truth. 
To guarantee the quality of the manual checking,
% the expertise of 
one author and two developers joined to label the data, and each of them possess
% each possessing 
over five years of software development experience.
% Their meticulous approach to labeling ensures a high quality of accuracy. 
% To prevent discrepancies, 
The two developers independently labeled the samples for the presence of vulnerabilities. 
% In the event of a disagreement, 
For the disagreement, the author intervened as a mediator to achieve a consensus.
% This process effectively eliminated any mislabeled samples. 
The manual checking
% results 
shows that the \tool finds 23 mislabeled samples in the dataset. 
% We believe that these mislabeled samples (also known as noisy samples) prevent the model from learning the correct vulnerability structure to some extent.

We broadly classify these mislabeled samples into two categories: 4(2\%) null functions and 19(9.5\%) unimplemented functions. Figure~\ref{disscussion}(B) shows a sample obtained in the FFMPeg+Qemu dataset with an empty function for this code sample. This sample does not have any syntax errors, but is mislabeled as positive. 
% However, it is not functional after the developer commits the new code, which the dataset labels as a function of vulnerability. 
% This is clearly an inappropriate labeling mechanism. 
Figure~\ref{disscussion}(C) shows an example of the unimplemented function, which does not involve any vulnerability but is labeled as positive in the dataset. The case analysis shows that \tool is able to well leverage the mislabeled data for more accurate vulnerability detection.
% another case where we \yun{can observe that } believe that it does an incomplete function, but the code itself does not contain a vulnerability. 
% We believe this case is a form of incorrect labeling.

\subsection{Comparison with Trovon}

The recent approach Trovon~\cite{DBLP:journals/ese/GargDJCPT22/Trovon} also aims at mitigating the data noise issue in vulnerability detection. It trains a Seq2seq model~\cite{DBLP:journals/corr/AbadiABBCCCDDDG16/seq2seq} based on the code fragment pairs (i.e., the pairs of vulnerable and fixed samples). Our proposed \tool is essentially different from Trovon in the following aspects:

\textbf{(1) Training data.} The approaches of Trovon~\cite{DBLP:journals/ese/GargDJCPT22/Trovon} and \tool exhibit significant disparities in the training data.
% Both approaches center around samples containing vulnerabilities, but 
Trovon utilizes a training set that comprises both before-fix (vulnerable) and after-fix (non-vulnerable) pairs to construct a classifier, aiming at investigating the
% . The primary aim is to thoroughly investigate the 
distinctions between these two sets. In contrast, \tool only utilizes the vulnerable samples as the training data based on the assumption that non-vulnerable samples potentially contain noisy data~\cite{DBLP:journals/tse/CroftXB23/311, DBLP:conf/msr/CroftBC22/312}.
% adopts a different perspective, regarding non-vulnerable samples as potentially noisy data. \tool solely relies on high-quality vulnerable samples as the training data and assigns greater significance. 
Remarkably, \tool pioneers the positive and unlabeled learning problem in the software vulnerability detection field. The approach involves a distance-aware label selection module and relies on
% for relying on 
high-quality vulnerable samples in the training process. Furthermore, \tool incorporates a mixed-supervision representation learning module to continuously train classifiers. These processes help mitigate the impact of noisy labels in vulnerability detection.

\textbf{(2) Handling of the Noise Issue.}
% Issue of noise.} 
Trovon and \tool address the noise issue in different ways. Specifically,
% are designed to address distinct types of noise. 
Trovon adopts a methodology to exclusively employ
% where it exclusively employs 
after-fix (non-vulnerable) and before-fix (vulnerable) samples for
% while disregarding unchanged ones, with the primary goal of 
minimizing noise within the training dataset. However,
% it is essential to acknowledge that, 
even after the fixing process, latent vulnerabilities may still persist in the data. This was also highlighted in Croft's work~\cite{DBLP:journals/tse/CroftXB23/311}, which revealed that non-vulnerable labels tend to be of subpar quality. 
% As a response to these challenges, \tool takes a different approach. It exclusively utilizes samples containing vulnerabilities for training purposes. 
Therefore, \tool evaluates the quality of non-vulnerable labels before usage. Specifically,
% To evaluate the quality of non-vulnerable labels, 
\tool incorporates a distance-aware label selection module, which generates pseudo-labels to assist in evaluating the quality of non-vulnerable labels.

\textbf{(3) Treatment of the Data Scarcity Issue.}
Another noteworthy aspect of \tool is its ability to achieve high performance with only a small amount of labeled data. Trovon relies on 
% knowledge acquired from observed data and necessitates 
a large amount of labeled data to build the classifier. Conversely, \tool utilizes a semi-supervised learning method that demands only a limited number of labeled data. 
% To address the challenge of data scarcity, 
\tool involves an inter-class distance prototype component derived from a small part of high-quality positive data. Subsequently, it involves a progressive fine-tuning process for further learning.

In summary, the proposed \tool is novel in its methodology design, and significantly different from Trovon in the
% distinctions between \tool and Trovon are significant, encompassing innovation, motivation, and methodology. Particularly, these differences are evident in 
training data, handling of the noise issue, and treatment of the data scarcity issue.

\subsection{Threats and Limitations}

One threat to validity comes from the dataset we construct. Following the positive and unlabeled learning setting, we construct a dataset with unknown labels on the existing dataset using only a portion of the positive labels. However, we do not use external unknown data. In the future, we will further collect a larger benchmark for evaluation.
% , obtaining better results on the current dataset.

The second threat to validity is the implementation of baselines. We try our best to replicate these weakly supervised in the PU setting based on the open-source code and the original paper to achieve the best experimental results.

Another validity to threat comes from the selection of positive samples. As we only use a part of the positive labeled samples, the selection of these samples affects the performance of \tool. We therefore repeatedly select these samples at random and take
% use 
the average
% of these cases 
as the experimental results.

%% file: Sections/7_Related_Work.tex
\subsection{Software Vulnerability Detection}
Software vulnerability detection is critical for ensuring software security by identifying and mitigating potential security risks. Nowadays, learning-based software vulnerability detection, as opposed to program analysis methods~\cite{redebug, symbolic, fuzz,argos}, has been shown to be more effective in identifying more types and numbers of vulnerabilities.
%a variety of methods are used in detecting vulnerabilities based on different perspectives.
Learning-based vulnerability detection techniques can be broadly classified into two categories based on the representation of source code and the learning model utilized: sequence-based and graph-based methods.

Sequence-based vulnerability detection methods~\cite{DBLP:journals/corr/abs-1708-02368, 8322752, DBLP:conf/codaspy/GriecoGURFM16, app10051692, DBLP:conf/sigsoft/WangYGP0L22/finetune} convert code into token sequences. For example, VulDeepecker~\cite{vuldeepecker} uses code gadgets as the granularity to train a classifier using a bidirectional (Bi)-LSTM network. 
% \wxc{$\mu$VulDeepecker~\cite{DBLP:journals/tdsc/ZouWXLJ21} matches vulnerability rules to generate code attention. }
Russell \et~\cite{russell} utilize Convolutional Neural Networks (CNNs) and Recurrent Neural Networks (RNNs) to fuse different features from function-level source code. 
SySeVR~\cite{sysevr} extracts code gadgets by traversing AST generated from code and also 
%using 
{uses} a Bi-LSTM network.
% In addition, the pre-trained model CodeBERT~\cite{DBLP:conf/emnlp/FengGTDFGS0LJZ20/codebert} is also applied to vulnerability detection by fine-tuning. %LineVul~\cite{DBLP:conf/msr/FuT22} proposes a row-level vulnerability detection scheme based on the BERT architecture, allowing for more granular detection.

% Sequence-based vulnerability detection methods convert code into token sequences on the source code.
% VulDeepecker divides the code into code gadgets and uses this as the granularity to train the classifier using a Bi-LSTM network. $\mu$VulDeepecker~\cite{DBLP:journals/tdsc/ZouWXLJ21} generates code attention by matching vulnerability rules on top of the former. Rebecca L. Russell et al developed a code vulnerability detection tool at the function granularity level using CNN and RNN to fuse near and far features. The pre-trained models CodeBert and CodeT5 are able to extract structural, syntactic and contextual information from the code and apply it in a fine-tuned way to the vulnerability detection task. LineVul~\cite{DBLP:conf/msr/FuT22} proposes a row-level vulnerability detection scheme based on the BERT architecture, which allows for more granular detection.

Graph-based methods~\cite{DBLP:conf/icse/WuZD0X022, DBLP:conf/ijcai/DuanWJRLYW19, DBLP:journals/tifs/WangYTTHFFBW21, DBLP:conf/ictai/PhanNB17, DBLP:journals/infsof/CaoSBWL21,DBLP:conf/icse/WenCGZZL23/graphsim,DBLP:journals/corr/abs-2212-14274/meta-path} represent software code as graphs and use Graph Neutral Networks (GNNs) for software vulnerability detection. CPGVA~\cite{DBLP:conf/icait/WangZWXH18} combines the AST, CFG, and DFG and generates the code property graph (CPG) to vulnerability detection.
% , making the representation more rich and complex. 
Devign adds Natural Code Sequence (NCS) to the CPG and leverages the Gated Graph Neutral Networks (GGNNs), which preserve the programming logic of the source code.  
% LineVD~\cite{linevd} uses the Graph Attention Transformer (GAT), which characterizes the data and control dependency information to vulnerability detection. 

However, all these existing vulnerability detection methods require a large amount of labeled positive and negative samples and fail to achieve good performance with a small labeled size. In this paper, we propose a positive and unlabeled learning model for vulnerability detection for the scenarios without enough labeled data.
\subsection{PU Learning}

% Positive and Unlabeled (PU) learning is a semi-supervised approach that trains a classifier using only positive and unlabeled samples. PU learning is beneficial in scenarios where negative samples are either challenging to acquire or expensive. There are three main categories of PU learning techniques: Two Step Techniques, Biased PU Learning, and Class Prior Incorporation.

Positive and Unlabeled (PU) learning setting is a weakly supervised learning scenario that aims to learn a classifier with only positive and unlabeled samples~\cite{DBLP:journals/ml/BekkerD20/pusurvey, DBLP:journals/ker/KhanM14, DBLP:journals/corr/ClaesenSGMM15, DBLP:conf/www/ZupancD18, DBLP:conf/fat/WuH22}. 
% PU learning is beneficial in practical applications~\cite{DBLP:journals/corr/ClaesenSGMM15, DBLP:conf/www/ZupancD18, DBLP:conf/fat/WuH22} where the number of labels is limited, or the data contains label noise. 
% We introduce the following two categories in PU setting, including biased PU learning and class-prior incorporation.
% The PU learning methods can be broadly classified into three categories, including two-step techniques, biased PU learning, and class-prior incorporation. 
% Two-step techniques involve two steps. First, it identifies negative samples from unlabeled samples through heuristic methods~\cite{}. For example, SPY~\cite{} selects samples among positive samples into unlabeled samples, which uses them to train the classifier to obtain the lowest threshold of the spy sample distribution. 
% 1-DNF~\cite{} constructs a feature set and identifies negative features by the frequency of occurrences. 
% %Rocchio~\cite{} determines the sample's label by the proximity of the samples to positive samples. 
% Second, the classifier is trained based on positive and negative labels, which further selects the reliable samples and the best classifier. For example, \wxc{need to add Roc-SVM..., S-EM, and CR-EM.}
An effective approach ~\cite{DBLP:conf/isnn/KeYZTLJ12, DBLP:journals/apin/KeJLZH18, DBLP:conf/icml/HsiehND15} considers unlabeled samples as negative samples with noise, and it is common practice to assign small weights to them. 
% As noise can cause misclassification, Biased PU learning often puts a higher penalty on misclassified positive samples and then adjust the model parameters. 
For instance, Lee \et~\cite{DBLP:conf/icml/LeeL03} use weighted logistic regression to learn from positive and unlabeled examples. %Liu \et~\cite{DBLP:conf/icdm/LiuDLLY03} propose an SVM-based biased formula approach. 
Another popular approach~\cite{DBLP:conf/aistats/Scott15, DBLP:journals/corr/JainWTR16, DBLP:conf/nips/JainWR16} involves incorporating knowledge of class priors into the training process, which adjusts the decision threshold of the classifier based on the estimated class prior probabilities.
% Class priors can be determined from domain knowledge, smaller fully labeled datasets, or directly estimated from labeled and unlabeled data. 
% For example, 
% POSC 4.5~\cite{DBLP:journals/tcs/DenisGL05} requires calculating positive and negative samples in each division of the set. 
Ward \et~\cite{ward2009presence} propose an expectation-maximization algorithm by class prior. 
Bekker \et~\cite{DBLP:conf/aaai/BekkerD18} propose a decision tree induction method for straightforward and efficient class prior estimation.

% Class Prior Incorporation integrates knowledge of class priors into the learning process. Specifically, it involves adjusting the decision threshold of the classifier based on the estimated class prior probabilities. In the Class Prior Incorporation, label frequencies can be obtained from domain knowledge, smaller fully labelled datasets estimated or directly from labelled and unlabelled data estimation. POSC 4.5 ~\cite{DBLP:journals/tcs/DenisGL05} needs to calculate positive and negative samples in each division of the tree. Ward at al. propose an expectation–maximization algorithm which needs class prior to train and rebalance the model ~\cite{ward2009presence}. ~\cite{DBLP:conf/aaai/BekkerD18} proposes a method based on decision tree induction for simple and efficient estimation of class priors.

In this paper, we propose a novel PU setting on software vulnerability detection. We also propose a distance-aware label selection module for correctly providing pseudo-label and a mixed-supervision representation learning module for better alleviating the influence of noise.

%% file: Sections/8_Conclusion.tex
% This paper proposes \tool, a positive-unlabeled learning model for vulnerability detection.
% It consists of a distance-aware label selection, which identifies reliable negative examples and provides pseudo-labeled samples. We also propose a representation and pseudo-label metric learning module to enhance the discriminative power of the vulnerability representation.
% Compared with the state-of-the-art methods, the experimental results on three popular datasets validate the effectiveness of \tool in PU and supervised settings. 
This paper focuses on the positive and unlabeled learning problem for vulnerability detection and proposes a novel model, named \tool. 
%a positive-unlabeled learning model for vulnerability detection.
\tool consists of a distance-aware label selection for generating pseudo-labels and a mixed-supervision representation learning module to alleviate the influence of noise.
Compared with the state-of-the-art methods, the experimental results on three popular datasets validate the effectiveness of \tool in PU and supervised settings.
In future, we will further collect a larger benchmark and obtain more data for vulnerability detection.

% , and the ablation studies further confirm the advantages of \tool. We further find the mislabeled samples in the dataset by the \tool.
%\section*{Data availability}
Our source code as well as experimental data are available at: \textit{{\http}}.